\documentclass[a4paper]{jpconf}
\usepackage{graphicx,amssymb,amsmath}

\begin{document}
\title{Decay of Resonaces in Strong Magnetic Field}

\author{Peter Filip}

\address{Institute of Physics, Slovak Academy of Sciences, D\'ubravsk\'a cesta 9, Bratislava $82103,\!$ Slovakia}

\ead{peter.filip@savba.sk}

\begin{abstract}
We suggest that decay properties (branching ratios) of hadronic resonances may become modified
in strong external magnetic field. The behavior of  $K^{\pm *}\!$, $K^{0*}$ vector mesons as well as 
$\Lambda^*(1520)$ and $\Xi^{0*}$ baryonic states is considered in static fields $10^{13}$-\,$10^{15}$ T.
In particular, $n=0$ Landau level energy increase of charged particles
in the external magnetic field, and the interaction of hadron magnetic
moments with the field is taken into account.
We suggest that enhanced yield of dileptons and photons from $\rho^0(770)$ mesons may occur if strong decay 
channel $\rho^0 \rightarrow \pi^+\pi^-$ is significantly suppressed. 
CP - violating $\pi^+\pi^-$ decays of pseudoscalar $\eta_c$ and $\eta(547)$ mesons in the magnetic field are discussed, 
and superpositions of 
quarkonium states $\eta_{c,b}$ and $\chi_{c,b}(nP)$ with 
$\Psi(nS), \Upsilon(nS)$ mesons in the external field are considered.
\end{abstract}

\section{Introduction} 
\label{sec:intro}
Modification of $\rho(770)$ decay properties in the magnetic field has been considered 
in relation to the possibility of $\rho^\pm$ meson condensation \cite{Chernodub} in sufficiently strong static magnetic fields. 
The paper \cite{Chernodub} suggests that the lowest energy of charged decay products ($\pi^+\pi^-$)
 is expected to exceed the mass of $\rho^0$ meson at field strength $eB \approx 0.14$ GeV$^2$.
This means $\rho^0 \rightarrow \pi^+\pi^-$ decays become suppressed and thus $\rho^0$ meson can be 
"stabilized" in the magnetic field background \cite{Chernodub}. 

One may immediately
consider the possibility that strong decays of other hadrons
can also be influenced in the magnetic field. Modification of decay probability into certain decay channels 
gives different branching ratios (BR) in the magnetic field when compared to the vacuum case. 
This may potentially affect reconstructed yields of resonances in heavy ion collisions, where
very strong magnetic fields are expected to be formed for a very short time \cite{B_HIC}, 
and reconstructed yields of the observed hadronic resonance can thus become decay-channel-dependent. 

In Sections \ref{sec:Rho}, \ref{sec:dileptS}, \ref{sec:Kmes} and \ref{sec:LambdaXi0}
of this contribution we investigate to what extent $\rho^0(770)$, $K^*(892), \Lambda^*(1520)$ and $\Xi^{0*}(1535)$ 
resonances may be affected by magnetic fields of strength $10^{13}$-\,$10^{15}$~T,
which may be present  
in collisions  of heavy nuclei \cite{B_HIC} at RHIC and LHC. 

Because decay $\rho^0 \rightarrow \pi^+\pi^-$ is the only possible strong decay channel of $\rho^0$ meson,
the enhancement of dilepton and photon production from $\rho^0$ decays can occur if $\rho^0 \rightarrow \pi^+\pi^-$
decay is closed for whatever reason.
In Section \ref{sec:dileptS} we speculate that 
excess of photons  and $e^+e^-$ or $\mu^+\mu^-$ pairs from $\rho^0$ decays may be expected
in the magnetic field of sufficient strength. 

Decay properties of ($q\bar q$) bound states in the magnetic field may be influenced also via different
mechanism, which involves quantum superposition of $J=0$ singlet and $L=1$ states with $J=1$
($m_z=0$ and $m_z=\pm1$) triplet states observed in Positronium \cite{positronium_WU, LLNL_StarkZeeman}. 
If such phenomenon happens in the case of hadrons \cite{myCPOD}, 
it may lead also to the enhancement of CP - violating decays $\eta \rightarrow \pi^+\pi^-$
in the magnetic field. This interesting behavior is discussed in Sections \ref{sec:CP_eta} and \ref{sec:Chi}.

\section{Energy of charged particles in strong magnetic field} 
\label{sec:LandauEnergy}
\begin{figure} [h,t]    
\begin{center} 
\includegraphics[width=34pc]{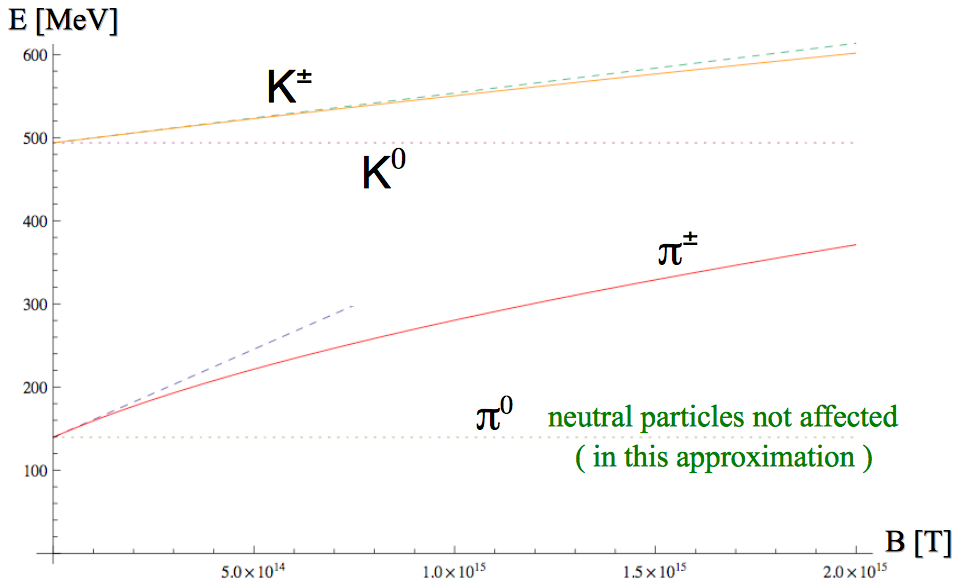} %%   % for LaTeX 2e
\vspace{-2mm} \caption{Energy of 
$\pi^\pm$ and $K^\pm$ mesons (with $p_z = 0$) in the magnetic field.} 
\label{Fig1_Landau}
\end{center}
\end{figure}

In the magnetic field, charged particle energy becomes quantized
and for the lowest possible $n=0$ Landau level \cite{Landau} the energy of electrons or positrons can be expressed as 
\begin{equation}
E[B,s_z] = \sqrt{m^2 + p_z^2 + |Q|B - 2QBs_z}\, + \Delta E[B,s_z]  
\label{Landau_Energy}
\end{equation}
where $m, Q$ are mass and charge of the particle, $p_z$ is 
momentum parallel to $\vec B$ field, and $\Delta E$ is a correction for the anomalous part of magnetic moment (see below).
For $p_z^2\ll m^2$, $Q=-e$ and $eB \ll m^2$ (assuming here $\Delta E \approx 0$) 
one obtains  ($e$ is electron charge magnitude)
\begin{equation}
E[B,s_z] \simeq m + p_z^2/2m + eB/2m + eBs_z/m\, .
\label{Landau_Energy_Approx}
\end{equation}
Term $eBs_z/m$ on the right side of Eq.(\ref{Landau_Energy_Approx}) corresponds to the energy of magnetic dipole
in external field $E = - \vec\mu \cdot\vec B$ with Dirac moment value $\mu = s_zQ/m$, and term $eB/2m$ 
accounts for the lowest Landau level \cite{Landau}. In Figure \ref{Fig1_Landau} we show how the energy of 
static ($p_z=0$) charged pions and kaons increases in magnetic field due to $n=0$ Landau level energy 
according to Eq.(\ref{Landau_Energy}). Dashed lines are obtained using Equation (\ref{Landau_Energy_Approx})
and dotted lines show the behavior of neutral ($J=0$) mesons.

One can also read from Eq.(\ref{Landau_Energy}) that 
energy of electrons with $s_z = -1/2$ and positrons with $s_z = +1/2$ does not change 
with magnetic field (if one neglects anomalous $\approx\!10^{-3}$
part of their magnetic moments). 

Particles with internal structure have gyromagnetic ratio $g$ substantially different from Dirac value ($g = 2$)
and "anomalous" magnetic moment needs to be taken into account properly. 
Energy dispersion relation for neutron in the magnetic field \cite{ASTRO} is
\begin{equation}
E[B,s_z] = \Big[p_z^2 + \Big(\sqrt{m_n^2 +p_T^2} - (2 s_z) \kappa_n \mu_N B \Big)^2 \Big]^{1/2}
\label{Eq:neutron}
\end{equation}
where $\kappa_n = -1.91$ 
is neutron magnetic moment in units $\mu_N = e\hbar /2m_p = 3.15\cdot 10^{-14}$ MeV/T. 
For small magnetic fields: $E[B,s_z] \approx m_n + (2 s_z) |\kappa_n|\mu_N B $ is obtained from Eq.(\ref{Eq:neutron}) using
($p_z, p_T \approx 0$), which corresponds to static magnetic dipole interaction $E = -\vec \mu_n \cdot \vec B$ . 
We shall use Eq.(\ref{Eq:neutron}) also for $\Lambda$ and $\Xi^0$ baryons with their 
magnetic moments $\mu_{\Lambda^0} = -0.61\mu_N$ and $\mu_{\Xi^0} = -1.25\mu_N$. 
For neutral $J=3/2$ particles $\Lambda^*(1520)$ and $\Xi^{0*}(1530)$, 
term $(2 s_z)$ in Eq.(\ref{Eq:neutron}) can replaced by $(2s_z/3)$ for obtaining the energy of $s_z = \pm 3/2$ substates. 

Dispersion relation for protons in the magnetic field \cite{ASTRO} can be expressed for $n=0$ Landau level as 
\begin{equation}
E[B,s_z] = \Big[p_z^2 +m_p^2 \Big(\sqrt{1+(1-2s_z)B/B_c^p} - s_z \kappa_p B/B_c^p \Big)^{\!2\,}\Big]^{1/2}
\label{Eq:proton}
\end{equation}
where $B_c^p$ constant is $m_p^2 c^2/e\hbar = 1.48\cdot 10^{16}$ T and $\kappa_p = 1.79$ is anomalous part
of proton magnetic moment.  
In weak magnetic fields ($B \ll B_c^p$) one gets 
$E[B,s_z] \approx m_p + eB/2m_p - (2s_z)[1+\kappa_p] \mu_N B$ for $p_z \approx 0$, which
corresponds well also to Eq.(\ref{Landau_Energy_Approx}) using $\mu_p = (1+\kappa_p)\mu_N$.

Behavior of electrons in very strong magnetic fields was calculated by J.~Schwinger, and for $B > 10^{11}$ T the 
correction term $\Delta E$ in Eq.(\ref{Landau_Energy})  
is \cite{ASTRO}
\begin{equation}
\Delta E[B,s_z]= 2s_z m_e (\alpha/4\pi)[\ln(2B/B_c^e) - (C+3/2)]^2 + \dots 
\label{Eq:Leptons}
\end{equation}
where $\alpha \simeq 1/137$, $B_c^e = m_e^2c^2/e\hbar = 4.414\cdot 10^{9}$~T, 
and $C=0.577$ is Euler's constant. 
For fields $B < B_c^e$, the correction for anomalous magnetic moment  
is $\Delta E =  2s_z(\alpha/4\pi)m_e B/B_c^e $, which corresponds to 
$\Delta E \approx  2s_z 10^{-3}|\mu_e| B$. For muon one can replace $m_e \rightarrow m_\mu$ in the equations,
which gives $B_c^\mu = B_c^e  (207)^2 = 1.89\cdot 10^{14}$ T, 
and for $B\ll B_c^\mu$ we have $\Delta E \approx  2s_z 10^{-3}|\mu_\mu| B$.

\begin{figure} [h,t]    
\begin{center} 
\includegraphics[width=32pc]{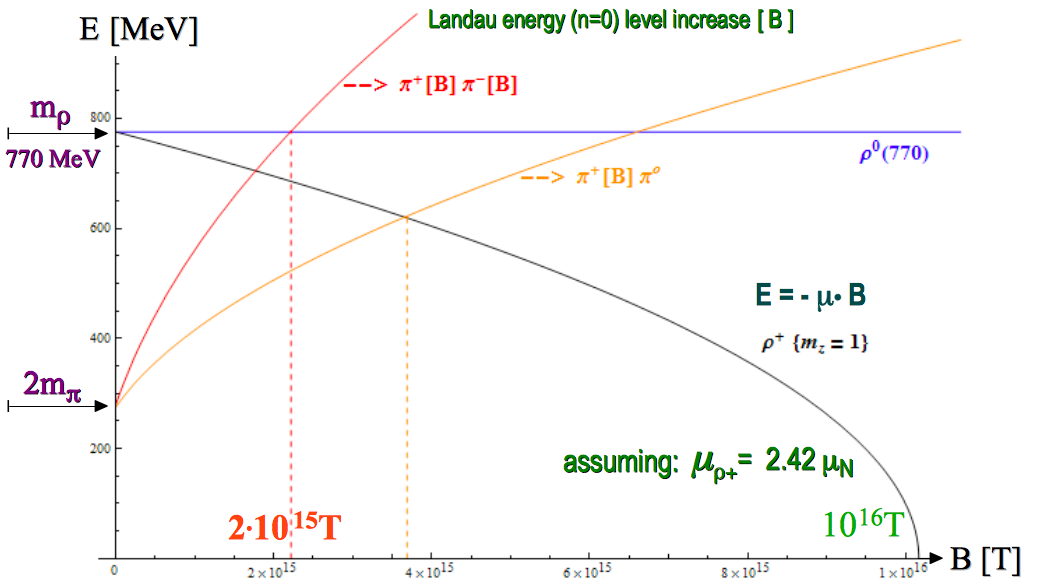} 
\vspace{-2mm} \caption{Energy of $\rho^0$ and 
$\rho^+$ ($s_z = +1$) mesons \cite{Chernodub} and $\pi^+\pi^{-(0)}$ pairs in magnetic field.} 
\label{Fig2rho}
\end{center}
\end{figure}

\section{Strong decays of $\rho(770)$ mesons in the magnetic field} 
\label{sec:Rho}
Let us summarize the mechanism \cite{Chernodub} which allows the lifetime of $\rho(770)$ 
mesons to be prolonged 
due to the suppression of  $\rho \rightarrow \pi\pi$ strong decays. 
Energy of charged pseudoscalar ($s = 0$) mesons increases in magnetic field according to Eq.(\ref{Landau_Energy}) 
and therefore at some critical field value $B^{cr}$ the mass of decay products may become equal to
the mass of decaying neutral hadron.
For $\rho^0(770)$ meson this occurs in field $B^{cr}\approx 2.2\cdot 10^{15}$~T, 
as shown in Figure \ref{Fig2rho}. Decay 
$\rho^0 \rightarrow \pi^+\pi^-$ becomes therefore energetically forbidden in fields $B >B^{cr}$.

Magnetic moment $\mu_{\!\rho^0\!}$ of $\rho^0$ meson $(u\bar u - d\bar d)/\sqrt{2}\,$  state  
is usually \cite{Chernodub} considered to be zero, neglecting in this way quadratic Zeeman interaction \cite{myCPOD}
of $\rho^0(s_z\!=\!0)$ substate  
(see the behavior of ortho-Positronium $(e^+e^-)$ $J=1$ state in magnetic field \cite{PositroniumAnn}).
However, charged $\rho^+ = (u\bar d)$ and 
$\rho^- = (d\bar u)$ mesons should have anomalous magnetic moments, which can be estimated 
also from the simple constituent quark model. 
For $u,\bar d$ quarks with parallel spins in $\rho^+$($J=1$) meson ground state, one has
$\mu_{\rho^+} =  (\mu_u + \mu_{\bar d}) = 1.85\mu_N+0.97\mu_N = \pm 2.82\mu_N$.

In Figure \ref{Fig2rho} we show energy decrease of $s_z = +1$ spin projection of charged $\rho^+$ mesons
evaluated according to publication \cite{Chernodub}, where
condensation of $\rho^\pm$ mesons is suggested to occur in static magnetic field. 
For majority of our estimates, however, it will be sufficient to consider weak-field approximations 
of the exact energy dispersion relations presented in Section \ref{sec:LandauEnergy}.

\section{Enhanced $\gamma$ and $l^+l^-$ production from $\rho^0$ meson decays}
\label{sec:dileptS}
In the case of $\rho^0\rightarrow \pi^+\pi^-$ decay suppression, 
the only potentially feasible strong decay channel $\rho^0 \rightarrow \pi^0\pi^0$
is forbidden by $C$ parity and isospin conservation. 
Therefore, closing of $\pi^+\pi^-$ decay channel ($BR=99\%$ in vacuum) has
intriguing consequences: other decay channels of type $\rho^0 \rightarrow \pi^0\gamma$ ($BR=6\cdot 10^{-4}$)
and $\rho^0 \rightarrow \eta\gamma$ ($BR=3\cdot 10^{-4}$) 
may become strongly enhanced as shown in Table \ref{tab:rho}. 
This  means decays of $\rho^0$ meson in sufficiently strong magnetic field will produce excessive photons 
(from channels containing $\pi^0$, $\eta$) and possibly also dilepton pairs. 

\begin{center}
\vspace{-7pt} 
\begin{table}[h]
\caption{\label{tab:rho}$\rho^0$ decay branching ratios (BR) in B field.} 
\centering
\begin{tabular}{c|c|c} 
\br
 channel & BR in $B\approx 0$ & BR in $B>2$$\cdot$$10^{15}$~T \\ 
\mr
$\pi^+\pi^-$ &  99\% & 0 \\
$\pi^+\pi^-\gamma$ &  0.9\% & 0 \\
$\pi^0\pi^0$ & 0 & \,\,\,0$^*$ \\ 
$\pi^0\gamma$ & $6\cdot 10^{-4}$ & 64\% \\  %% 60.5\% \\
$\eta\gamma$ & $3\cdot 10^{-4}$ & 31\%  \\%% 30.3\% \\
$l^+l^-$ & $0.9\cdot 10^{-4}$ & \raisebox{1pt}{$\scriptstyle\lesssim$}\,\,$\!1$\%$^{\,\,}$ \\
$\pi^0\pi^0\gamma$ & $0.4\cdot 10^{-4}$ & \, 4\% \\ %% \,4.5\% \\
$\pi^+\pi^-\pi^0$ &  $1\cdot 10^{-4}$ & 0 \\
$\pi^+\pi^-\pi^+\pi^-$ &  $0.2\cdot 10^{-4}$ & 0 \\
\br
\end{tabular}
\end{table}
\vspace{-17pt} \small{$^*$Assuming $\rho^0\!\rightarrow \pi^0\pi^0$ remains closed in the magnetic field.\,\,}
\end{center}
\vspace{10pt}

Regarding dilepton pair production via $\rho^0 \rightarrow l^+\l^-$ decays in the strong magnetic field,
one can make the following simplified consideration:
$J=1$ angular momentum of decaying $\rho^0$ meson must be conserved during decay process 
and this is achieved by parallel spin orientation (e.g. $\uparrow\!\uparrow $ or $\downarrow\!\downarrow $
for $J_z = \pm 1$) of
produced leptons $l^+l^-$. 
From Eq.(\ref{Landau_Energy}) it is then clear that energy $E[B,s_z]$
of one of the leptons will be increasing with the magnetic field, while energy of the other lepton 
remains almost constant (corrections $\Delta E$ from Eq.(\ref{Eq:Leptons}) are negligible).  At field $B = 2.5\cdot 10^{15}$~T
the Equation (\ref{Landau_Energy}) gives $E[B,s_z=+1/2] \approx 550$~MeV  for electrons or muons,
which  allows $\rho^0 \rightarrow l^+l^-$ decays to happen with restricted phase-space
(energy of $\,l^+l^-$ pair does not exceed $\rho^0$ mass). 

Directions of the momenta of produced charged leptons leaving the region
of overcritical field $B^{cr}$ will be affected by the Lorentz force $F = q \cdot \vec v\times \vec B$, which distorts (influences) 
the reconstructed invariant mass of dilepton pairs. Therefore $d N/dM_{l^+l^-}$ distribution of dileptons
from the considered $\rho^0$ meson decays (in volume containing 
strong magnetic field $B \approx B^{cr}$) should/can
be different from the expected $d N/dM_{l^+l^-}\!$ spectrum in ($B=0$) vacuum case.

It is interesting to point out, that excessive elliptic flow asymmetry $v_2$ of photons 
has been observed by PHENIX collaboration \cite{PHENIX} in Au+Au collisions. A simulation with  
enhanced decay probability of $\rho^0\rightarrow \gamma+\pi^0$ and $\rho^0\rightarrow\gamma+\eta$ channels
could easily clarify, whether enlarged $v_2$ of photons observed \cite{PHENIX}
may be explained by suppressed (for whatever reason) $\rho^0 \rightarrow \pi^+\pi^-$ decays. 

Assuming, that e.g. only  3\% of $\rho^0$ mesons produced in heavy ion collision 
have  $\pi^+\pi^-$ decay channel closed, branching ratios from the right side of Table 1 can easily generate
the excess of photons or dileptons many times above the expected yield, without
noticeably lowering the amount of $\rho^0 \rightarrow \pi^+\pi^-$ decays from the remaining 
(97\% unaffected) $\rho^0$ mesons.

\section{The case of $K^{0*}$ and $K^{\pm*}$ meson decays} 
\label{sec:Kmes}
\begin{figure} [h,t]    
\begin{center} 
\includegraphics[width=32pc]{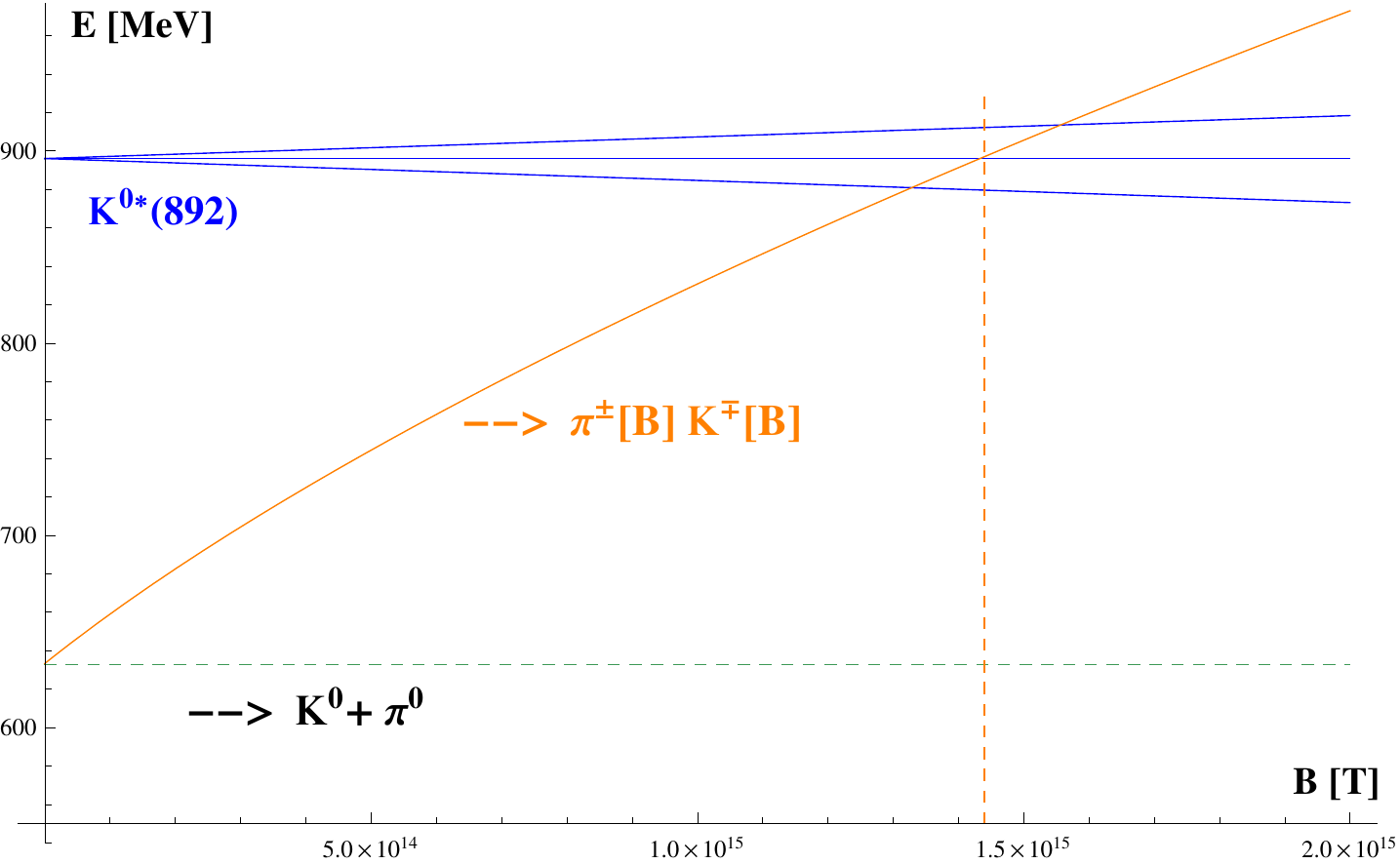} 
\vspace{-2mm} \caption{Energy of $K^{0*}(896)$ substates  $(m_z=\pm 1,0)$
and of $K+\pi$ decay products in $B$ field. } 
\label{Fig3K0}
\end{center}
\end{figure}

Mesons $K^{0*}(d\bar s)$ and $\bar K^{0*}(\bar d s)$ are ($J=1$) particles 
of mass $M= 896$~MeV, which decay via strong decay channels
$\bar K^{0*}$$(K^{0*})\rightarrow K^\pm + \pi^\mp$ and $K^{0*}\rightarrow K^0 + \pi^0$ with probabilities 66\% and 33\% 
(ratio 2:1 is determined by Clebsch-Gordan coefficients originating from the isospin conservation). 
In Figure \ref{Fig3K0} we show the energy of $(K^0\pi^0)$ and $(K^\pm\pi^\mp)$ decay products and of decaying $K^{0*}$
meson in the magnetic field. Triplet splitting observed for $K^{0*}$ state comes from the magnetic moment of
$K^{0*}$ meson, which can be estimated to be $\mu_{K^{0*}} = (\mu_d + \mu_{\bar s}) = - 0.36\mu_N$
assuming parallel spin orientation for $d,\bar s$ quarks in 1$S$ quantum state. Interaction energy of $(K^{0*})$ particle with
magnetic field is evaluated using expression $E[B , J_z] = m_{K^{0*}} - B J_z \mu_{K^{0*}}$
which agrees with 
Eq.(\ref{Eq:neutron}) in small field limit, when $(2s_z) \rightarrow J_z$ substitution is used and $p = 0$.

It is obvious that $K^{0*}\rightarrow K^- + \pi^+$ decay is kinematically forbidden in field $B > 1.5\cdot 10^{15}$~T
since the energy of $(K^-,\pi^+)$ decay products 
exceeds the mass of $K^{0*}$ meson. In such case the isospin conservation rule (leading to 2:1 ratio of
decay probabilities) becomes violated, exactly as it happens with $D^{0*} \rightarrow D\pi$ strong decays in vacuum
($B=0$). 

\begin{figure} [h,t]    
\begin{center} 
\includegraphics[width=32pc]{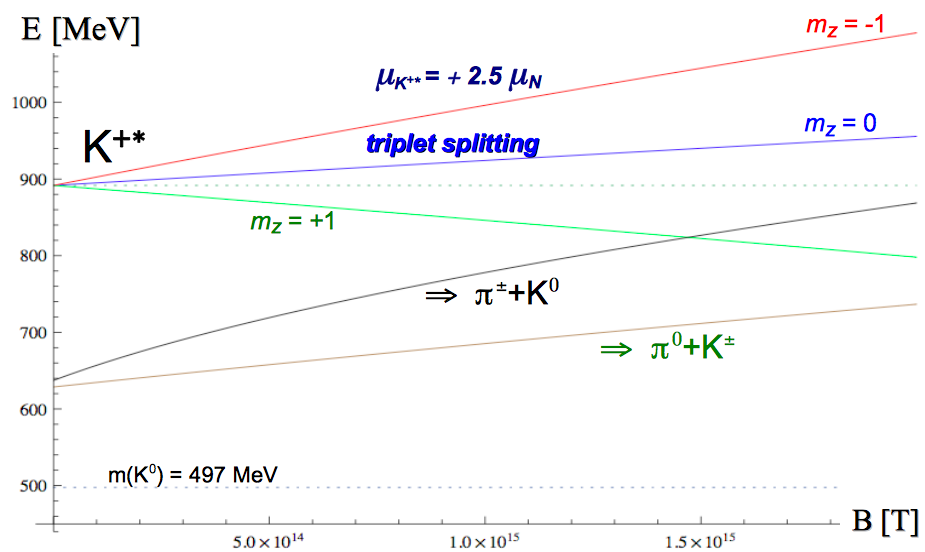} 
\vspace{-2mm} \caption{Energy of $K^{+*}(892)$ substates  $(\pm 1,0)$
and of $K+\pi$ decay products in $B$ field. } 
\label{Fig4Kpm}
\end{center}
\end{figure}

Let us have a look what happens in the case of charged $K^*$ meson decays. 
Magnetic moment of $K^{\pm*}$ mesons
$\mu_{K^{\pm *}} = \pm(\mu_u + \mu_{\bar s}) = \pm (1.85 + 0.61) = \pm 2.46\mu_N$ is 7x larger compared
to $\mu_{K^{0*}}$ case and the additional contribution from $n=0$ Landau level modifies the energy of
$K^{\pm*}$ mesons. In the "small" field limit and static ($p_z = 0$) case one can write 
\begin{equation}
E[B, J_z] = m_{K^{\!*}}  + |q_{K^{\!*}}| B /2m_{K^{\!*}} - J_z \mu_{K^{\!*}} B
\label{Landau_EnergyWxx}
\end{equation}
which is sufficient for our purposes here. From Figure \ref{Fig4Kpm} one can conclude that
both decay channels ($K^{+*} \rightarrow K^+ + \pi^0$ and $K^{+*} \rightarrow K^0 + \pi^+$) are affected and that
the isospin violation effects are going to be much weaker if compared to the case of $K^{0*}$ decays.

If $K^{0*}$ and $K^{\pm*}$ resonance yields in ultra-relativistic heavy ion collisions are determined from their
$K^\pm + \pi^\mp$ and $K^0_s + \pi^\pm$ decays, and strong magnetic field
effects we discuss here are really significant, one can expect that $K^{0*}$, $K^{\pm*}$ yields (determined with the
assumption of isospin conservation in strong decays) will be different. 
This happens because of different degree of the isospin
violation we predict here for $K^{0*}$ and $K^{\pm*}$ mesons in the external magnetic field. 

Additionally, from Figure \ref{Fig4Kpm} one can conclude that different $J_z$ spin projections of $K^{\pm*}(892)$
mesons are affected differently by the restricted phase-space available for $K_s^0 + \pi^\pm$ decays, which may
consequently lead to the measurable tensor polarization of $K^{\pm*}$ 
mesons. 

\section{Decay of $\Lambda^*$ and $\Xi^{0*}$ in strong magnetic field}
\label{sec:LambdaXi0}
Energy of neutral baryons in the magnetic field is described by Eq.(\ref{Eq:neutron}) which leads to
quadruplet energy splitting for $J=3/2$ case. Magnetic moment of $\Xi^{0*}(1530)$ resonance
can be estimated as $\mu_{\Xi^{0*}} = 2\mu_s + \mu_u = 0.62\mu_N$ 
using the constituent quark model for $u^\uparrow s^\uparrow s^\uparrow$ quarks 
with parallel spins in $L=0$ quantum state, while
for $\Lambda^*(1520)$ we shall use value
$\mu_{\Lambda^*} = -0.2 \mu_N$ \cite{Lambda1520}. 
 
\begin{figure} [h,t]    
\begin{center} 
\includegraphics[width=32pc]{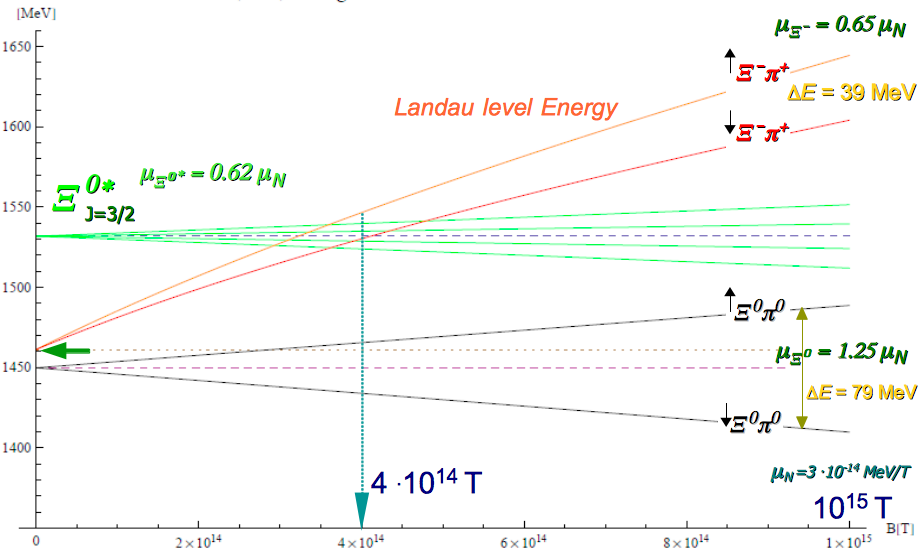} 
\vspace{-2mm} \caption{Energy of $\Xi^{0*}(1532)$ baryon 
and of its decay products $\Xi^{-}\pi^+$ and $\Xi^{0}\pi^0$ in static magnetic field
(using $\mu_{\Xi^{0*}}=2\mu_s+\mu_u = 0.62\mu_N$ and $s_z = \pm 1/2$ orientations 
for $\,\Xi^{-}$ and \, $\Xi^{0}$).} 
\label{Fig5_Xi0}
\end{center}
\end{figure}

Figure \ref{Fig5_Xi0} suggests that $\Xi^{0*} \rightarrow \Xi^- +\pi^+$ decay channel becomes energetically
forbidden in field $B > 4\cdot 10^{14}$~T for both orientations of $\Xi^-$ spin $s_z = \pm 1/2$, while
strong decays $\Xi^{0*} \rightarrow \Xi^0 +\pi^0$ remain kinematically allowed. At even higher field strengths
the interaction of magnetic moment $\mu_{\Xi^{0}} = 1.25\mu_N$ with the external field (for $s_z = - 1/2$) may bring
the energy of $\Xi^{0}+\pi^0$ pair above the mass of $\Xi^{0*}$ substates, but this is not our concern here.
Our main conclusion is, that due to the restricted phase space for $\Xi^{0*} \rightarrow \Xi^- +\pi^+$ decays
the isospin conservation in decays $\Xi^{0*} \rightarrow \Xi +\pi$ can become violated in a sufficiently
strong (static) magnetic fields. 

Lifetime of $\Xi^{0*}$ baryon state ($\approx 21$ fm/c) is considerably longer
compared to the expected magnetic field duration in relativistic heavy ion collisions, and therefore we are reluctant
to relate the smaller yield of $\Xi^{0*}$ baryons discussed at RHIC \cite{Xi0Xi0_RHIC}
with the magnetic field-induced phenomena we study here. However, lifetime $\tau_{\Lambda^*} \approx 13$ fm/c
of resonance $\Lambda^*(1520)$
is shorter, and therefore the behavior of  $\Lambda^*$ in the magnetic
field may be of some relevance.

\vspace{-3mm}

\begin{figure} [h,t]    
\begin{center} 
\includegraphics[width=32pc]{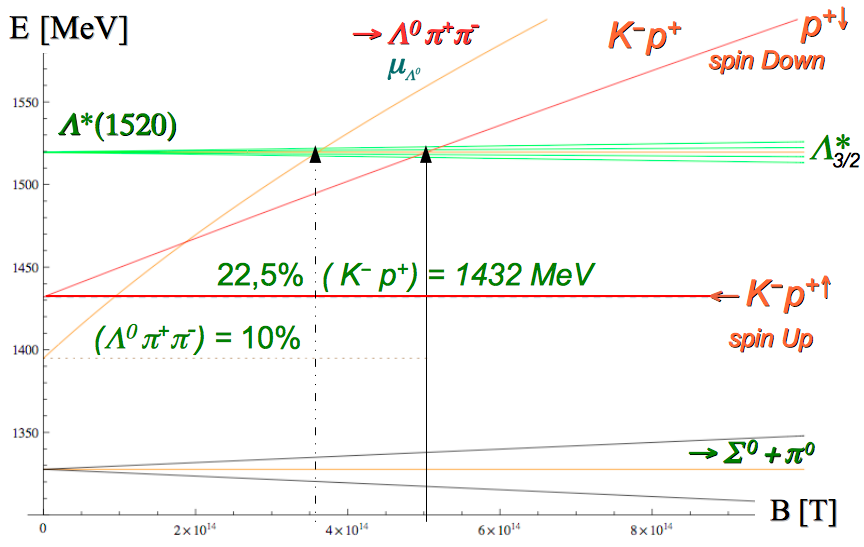}  
\vspace{-2mm} \caption{Energy of $\Lambda^{*}_{3/2}(1520)$ baryon 
$(s_z$=$\pm 1/2,$$\pm 3/2)$ substates  
(using $\mu_{\Lambda^{*}}$=$- 0.2\mu_N$)
and of its decay products $K^-p^+$, 
$\,\Lambda^0\pi^+\pi^-$ and $\Sigma^0\pi^0$ in the external static magnetic field.} 
\label{Fig6_Lambda}
\end{center}
\end{figure}

In Figure \ref{Fig6_Lambda} we show that decay channel $\Lambda^* \rightarrow \Lambda^0 \pi^+\pi^-$ becomes
kinematically closed in field $B > 3.5\cdot 10^{14}$~T, while channel  $\Lambda^* \rightarrow \Sigma^0 \pi^0$
remains open in even much higher magnetic fields. 

For decay $\Lambda^* \rightarrow \Lambda^0 \pi^+\pi^-$ we do not show $\Lambda^0$ energy
splitting due to $\mu_{\Lambda^0} = -0.613 \mu_N$ interaction with $B$ field, because the effect is small, and this channel
is not relevant here. However, for   $\Lambda^* \rightarrow p+ K^-$ decay channel, which is used
to reconstruct $\Lambda^*(1520)$ in the experiments, the interaction $\Delta E = - \vec B \cdot \vec \mu_p$ is important. 

For $s_z = -1/2$ proton spin projection, the energy of $(p+K^-)$ pair rises above $\Lambda^*$ 
mass at $B > 5\cdot 10^{14}$~T. This happens because $n=0$ Landau energy increase of $K^-$ meson
and Proton is supplemented by $\Delta E = + \mu_p B$ positive energy contribution. However,
in $p^\uparrow (s_z = +1/2)$ case, the interaction of $\mu_p = 2.79 \mu_N$ with the field just compensates
the rising contribution from $n=0$ Landau energy of charged $K$ and $p$ particles, what gives almost
horizontal line at $E = 1432$~MeV if Figure \ref{Fig6_Lambda}. Thus 50\% of $\Lambda^* \rightarrow p+ K^-$
decays become suppressed in $B > 5\!\cdot \!10^{14}$~T  field, while $\Lambda^* \rightarrow n+ \bar K^0$
channel remains open and (almost) unaffected.

Precise measurement of $\Lambda^*(1520)$ production in heavy ion collisions 
might be interesting, because $\Lambda^*$ signal observed in peripheral and central 
Au+Au collisions at RHIC \cite{STAR_Lambda_star} 
is missing in non-central Au+Au collisions \cite{Lambda_STAR_SUBATECH}. 
It remains to be resolved whether rescattering mechanism or just statistics, or
decay channel suppression phenomenona discussed here are responsible for the unobserved $\Lambda^*(1520)$ peak
in non-central Au+Au collisions at RHIC.
 
Decays of $\Delta(1232)$ resonance  with its short lifetime $\tau \approx 2$ fm/c  may also be
influenced by magnetic fields of strength $B\! \approx\! 5\!\cdot\! 10^{14}$~T. Interested readers may
find related information in \cite{myPoS_Baldin}.

\section{Enhanced CP violation in $\eta_c$ and $\eta(547)$ decays due to magnetic field}
\label{sec:CP_eta}
In the preceding sections we have studied the interaction of magnetic field with charge and magnetic moments
of hadrons. We have found that isospin conservation rules in strong decays of resonances
may become violated due to kinematical (energy conservation) and phase-space reasons.
However, the interaction of constituent quark magnetic moments with external magnetic field may change
also the internal structure of hadrons and their decay properties, as we shall describe here.

It has been mentioned already by Gell-Mann and Pais \cite{GellMan} that rigorous conservation
of C parity should be expected only in the absence of external fields. Indeed, ortho-Positronium (o-Ps)
$s_z\,$=\,0 substate (which may decay only via o-Ps $\rightarrow 3\gamma, 5\gamma$ channels in vacuum)
obtains quantum admixture of para-Ps ($J=0$) state in the magnetic field. In such situation,
($e^+e^-$) bound state with $J=1, s_z = 0$ quantum numbers can decay into 
$\rightarrow \gamma\gamma$ pairs, which leads to experimentally observed \cite{WuPs}
"magnetic field quenching" of ortho-Ps $\rightarrow 3\gamma$ decays.
The phenomenon occurs because new eigenstates $\Psi_o^+$ and $\Psi_p^-$ 
of the Hamiltonian containing the interaction of $\mu_{e^+}$ and $\mu_{e^-}$ magnetic moments 
with external magnetic field are (for $B\ne 0$) superpositions
of the original C parity eigenstates: $\Psi_p (J=0) = ( \uparrow\downarrow - \downarrow\uparrow)/\sqrt{2}$ and 
$\Psi_o (J=1, s_z = 0) = ( \uparrow\downarrow + \downarrow\uparrow)/\sqrt{2}$\,:
\begin{equation}
\Psi_o^+ = \cos(\alpha) \Psi_o + \sin(\alpha) \Psi_p \qquad  \Psi_p^- = \cos(\alpha) \Psi_p - \sin(\alpha) \Psi_o
\label{eq:superposition}
\end{equation}

Mixing angle $\alpha$ depends on the external magnetic field $B$ as $\sin(\alpha)=y/\sqrt{1+y^2}$, where
$y=x/(1+\sqrt{1+x^2})$ and $x=4\mu_e B/\Delta E_{hf}$ (here $\Delta E_{hf}=8.4\cdot 10^{-4\,}$eV is hyperfine energy
splitting of Positronium \cite{PositroniumAnn}). Since constituent quarks have their magnetic moments (one may expect 
$\mu_c \approx 0.4\mu_N$ and $\mu_b \approx -0.07\mu_N$),  eigenstates $\Psi(nS)$ and $\Upsilon(nS)$ of $J=1$
Quarkonium  mesons should be affected \cite{myCPOD} by external magnetic fields in a similar way as ortho-Positronium. 
Hyperfine energy splitting between ($J=0$) $\eta_{c,b}$ mesons and ($J=1$) $\Psi, \Upsilon$ states is 116MeV 
and 71MeV, and the required magnetic field for the mixing effect to occur in Quarkonium 
is \cite{myCPOD}: $B\approx 10^{14} - 10^{15}$~T. This phenomenon (suggested at 2012 DSPIN conference \cite{myDSPIN})
allows mesons, in quantum superposition state given by Eq.(\ref{eq:superposition}), 
to decay via new channels (in magnetic field).
For example, $s_z = 0$ substate of $\Upsilon(1s)$ meson (decaying to $\rightarrow ggg$ channel in 82\% of cases) 
may disintegrate via much faster $\rightarrow 2g$ (gluons) channel in the magnetic field, which 
can shorten the lifetime of $\Upsilon$
mesons considerably \cite{myCPOD} and also suppress the amount of $\Upsilon \rightarrow l^+l^-$ decays.

However, in this contribution we would like to discuss another consequence of the above mentioned
quantum mixing phenomenon: In the magnetic field, para-Positronium $J=0$ state acquires admixture 
of o-Ps($J=1, s_z = 0)$ substate. This is not very interesting in Positronium case, because lifetime ratio
$\tau_{o-Ps} / \tau_{p-Ps} \approx 10^3$ results in very small influence \cite{PositroniumAnn} of the magnetic field 
on para-Ps decays (possibility to decay via $\rightarrow 3\gamma$ channel 
does not influence significantly quantum state which can decay via $\rightarrow 2\gamma$ "fast" channel by default). 

However, in the case of $\eta(547)$ and $\omega(782)$ mesons, the ratio
of lifetimes is opposite $\tau_\omega / \tau_\eta = 1.5\cdot 10^{-4}$, which means that $r=10^{-6}$ admixture of $\omega$ meson in $\Psi_p^- = \tilde \eta(547)$ superposition ($J=0$) state
introduces $\approx 0.6\%$ contamination from $\omega$  decays \cite{myMESON2014}. This allows $\tilde \eta(547)$
meson in the magnetic field to decay also into $\rightarrow \pi^+\pi^-$ final states, which is otherwise 
forbidden by CP conservation at the level BR($\eta \rightarrow \pi^+\pi^- \le 10^{-27}$) within Standard Model.
For $\eta_c - J/\Psi$ mixing, the enhancement of 
$\eta_{c} \rightarrow \pi^+\pi^-$ decays is much less significant ($\tau_{\Psi}/ \tau_{\eta_{c\,}}$$\gg$\,1). 

It seems that magnetic field is able to enhance CP - violating decays of bound hadronic systems via
quantum mixing - superposition phenomenon. This may occur only if "G" parity violation 
is significant ($J/\Psi $, $\omega(782)\!\rightarrow\! \pi^+\pi^-$ decays cannot
occur when G parity is conserved).

\section{Superposition of J/$\Psi$, $\Upsilon$ with $\chi_{c,b}$ states in the external field}
\label{sec:Chi}
When Quarkonium vector ($J\!=\!1$) mesons move in the magnetic field, they experience (in their rest frame)
electric field $E_t = -\gamma v_t B_z$, 
which can influence their internal quantum state in a specific way. For Positronium, such phenomenon
(motional Stark effect) has been studied carefully \cite{LLNL_StarkZeeman}. 
It has been established that $s_z = \pm 1$ states of ortho-Positronium acquire admixture from $1P$ ($e^+e^-$) states 
(these correspond to $\chi_c(J=0,1,2$) mesons in the case of Charmonium). 
For example, eigenstate $\tilde\Psi^{s_z\!}(c\bar c)$ with 
$J\!=\!1, L\!=\!0, s_z =\! +1$ quantum numbers of the Hamiltonian
containing the interaction term for external electric fields is 
\begin{equation}
\tilde\Psi^{+1}(2s) =\Psi^{+1}(2s) 
                          -  x_2 (\frac{\chi_{c2}^0}{\sqrt{6}} -\chi_{c2}^{+2}) 
                          -x_1 \frac{\chi_{c1}^0}{\sqrt{2}} 
                         +x_0 \frac{\chi_{c0}^0}{\sqrt{3}} 
\label{Stark_2s_m}
\end{equation}
where parameters $x_2, x_1, x_0$  depend on "hyperfine" energy splittings $\Delta E_{hf}$  between $\Psi(2s)$ and 
$\chi_{c0}, \chi_{c1},  \chi_{c2}$ mesons. One has 
$x_2 = 3\sqrt{2} \, q_c \langle r_0 \rangle E_t/\Delta E_{hf} (\Psi_{2s}\! - \chi_{c2}) = 
2\sqrt{2} \, e \langle r_0 \rangle E_t/130$MeV, where $\langle r_0 \rangle = 0.45$fm is the "size" of $\Psi(2s)$ state.
Expressions for $x_1$ and $x_0$ are the same as $x_2$, just $\Delta E_{hf}$ parameters are 175MeV and  271MeV
for $(\Psi_{2s}\! - \chi_{c1})$ and $(\Psi_{2s}\! - \chi_{c0})$ mass differences.

We mention this here to point out, that $\Psi(nS) - \chi_c(nP)$ and $\Upsilon(nS) - \chi_b(\tilde nP)$ 
superpositions (in external electric field)
are more significant in case of small $\Delta E_{hf}$ mass splittings between $\chi_{c,b}$ and triplet $J\!=\!1$ Quarkonium
states. From the known masses of Charmonium and Bottomium mesons
one can find that $\Psi(2s)$ and $\Upsilon(2s,3s)$ states are more affected by quantum mixing effects (in the 
magnetic and electric field) than $J/\Psi(1s)$ and $\Upsilon(1s)$ mesons. 

Admixtures of $\chi_{c,b}$ and $\eta_{c,b}$ states in $\Upsilon(ns)$ and $\Psi(ns)$ mesons
allow them to decay via "fast" ($\rightarrow gg$) channels, lowering the yield of $l^+\l^-$
pairs from their decays. We suggest that suppression/disappearance of 
$\Upsilon(2s,3s)$ yields observed in Pb+Pb collisions  at LHC \cite{CMS_uu} 
might possibly be related also to the quantum mixing behavior we have discussed here.

\section{Summary}
We have studied the influence of external magnetic and motional electric field on decay properties of mesons and baryonic
resonances. We suggest that isospin conservation rules determining branching ratios for various strong decay channels
of $K^*, \Xi^*, \Lambda^*$ hadronic resonances 
may become violated in the magnetic field ($B\approx 10^{14}$~T) 
due to the restricted phase-space and energy conservation. In the case of
$\rho^0 \rightarrow \pi^+\pi^-$ decay suppression, we speculate that enhanced photon and dilepton
production may occur. 
CP violation in $\eta \rightarrow \pi^+\pi^-$ decay channel has been predicted to be enlarged in the magnetic field,
and the influence of motional electric field on $\Psi(ns)$ and $\Upsilon(ns)$ Quarkonium states has been considered.

\ack 
This work was performed within project No.\,2/0197/14 of the Slovak Grant Agency~VEGA. 
Financial support from Slovak Research and Development Agency project No. APVV-0177-11 is acknowledged.
The author thanks to Slovak CERN committee for the encouraging support and to 
K.~\v Safa\v r\'ik for the kind hospitality at CERN, where some of ideas presented here had been elaborated. 
The invitation
to give a talk at WWND 2015 conference and excellent management of this scientific meeting
(by R. Bellwied, F. Geurts, and A. Timmins) are highly appreciated.

\end{document}